\documentclass[12pt,a4paper,float]{article}
\usepackage[latin1]{inputenc}
\usepackage{graphicx}
\usepackage{color}
\usepackage{amsmath}\usepackage{amssymb,bbm,bm,amsfonts,amssymb,mathrsfs,bbold}
\newcommand{\eq}[2]{\begin{align}\label{#1}#2\end{align}}

\newcommand{\nn}{\nonumber}
\newcommand{\pa}{\partial}
\newcommand{\ben}{\begin{enumerate}}\newcommand{\een}{\end{enumerate}}
\renewcommand{\Ref}[1]{(\ref{#1})}

\newcommand{\al}{\alpha}

\newcommand{\la}{\lambda}\newcommand{\ga}{\gamma}

\newcommand{\om}{\omega}

\newcommand{\sn}{{\rm sn}}
\newcommand{\cn}{{\rm cn}}
\newcommand{\me}{{ m_e}}

\begin{document}
\title{Vacuum energy for a scalar field with self-interaction in (1+1) dimensions}
\author{M. Bordag\thanks{bordag@uni-leipzig.de}\\
{\small Institute for Theoretical Physics, Universit{\"a}t Leipzig}\\
}
\date{\small 12.2.2021}
\maketitle

\begin{abstract}
We calculate the vacuum (Casimir) energy for a scalar field with $\phi^4$ self-interaction in (1+1) dimensions non perturbatively, i.e., in all orders of the self-interaction. We consider   massive and massless fields in a finite box with Dirichlet boundary conditions and on the whole axis as well. For strong coupling, the vacuum energy is negative indicating some instability.

\end{abstract}\thispagestyle{empty}

\section{\label{T1}Introduction}
The notion of {\it vacuum energy} results from the ground state of a quantum system, particle, or field. This is the state with no excitations. As known since early quantum mechanics, the energy of a harmonic oscillator,
\eq{1}{E_n&=\hbar\om\left(n+\frac12\right),
}
has a resudial amount,
\eq{2}{E_0&=\frac12\hbar\om,
}
which is called {\it zero-point energy}. Since in quantum field theory the groundstate is called vacuum state, or simply vacuum, this energy is called {\it vacuum energy}. The necessary sum over the degrees of freedom of a field,
\eq{3}{ E_0&=\frac12\sum_{(n)}\hbar\om_{(n)},
}
makes the vacuum energy, of course, infinite. While this was a problem in earlier times, by now powerful methods of regularization and renormalization are known, notably the zeta functional regularization and the renormalization assisted by the heat kernel expansion. The regularized vacuum  energy,
\eq{4}{ E_0(s) &= \frac{\mu^{2s}}{2}\sum_{(n)}\hbar\om_{(n)}^{1-2s},
}
with $s\to 0$ at the end, is finite and does exist, in distinction from \Ref{3}. The known price to be paid is the appearance of the arbitrary parameter $\mu$.

The simplest example known is a free scalar field in a box with Dirichlet boundary conditions. It has eigenfrequencies
\eq{5}{ \om_j &=\frac{\pi j}{L}
}
and its vacuum energy is
\eq{6}{  E_0 &= \frac{\mu^{2s}}{2} \hbar \sum_{j=1}^\infty \om_j^{1-2s}
	=\frac{\mu^{2s}}{2} \hbar \zeta_{\rm R}(2s-1)
	=-\frac{\pi\hbar}{24L}.
}
In this case, the continuation to $s=0$ is done by the Riemann zeta function and there was no pole term (which is, of course, not the general case). For a detailed explanation of the renormalization procedure, we refer to the appropriate literature, the book \cite{BKMM}, Sect. 4.3,  for instance.

A characteristic feature of the vacuum energy \Ref{6} is that it vanishes for $L\to\infty$. This is not surprising for two reasons. First, for dimensional reasons, and second, from the interpretation as the vacuum energy change brought in by the boundaries. Accordingly, there is a general expectation that the vacuum energy makes sense only relative to the empty space.

It must be mentioned that this remark touches on a principal question. Let's start from zero-point energy in quantum mechanics, say of one harmonic oscillator. Its ground state energy $E_0$ cannot be measured from experiments that were sensitive to a change in the energy levels, say by spectroscopy, but only by a change of the intrinsic frequency $\om$ of the oscillator. The first detection of zero-point energy this way is known from early quantum mechanics as a difference in vapor pressure between certain neon isotopes (see the contribution by Rechenberg in \cite{bord98}). This feature slipped over to quantum field theory.
What we can measure is always a change in energy, resulting in a force or pressure. The same holds also for the gravitational action. An attempt to give the vacuum energy of the empty space a physical meaning by introducing a cut-off, results in the known, too large by far, estimates for the cosmological constant (see, e.g., \cite{wein89-61-1}). Recently, this topic was discussed from a modern point of view in \cite{most19-11-314}.

So far it was not possible to attribute, in a meaningful way, vacuum energy to the empty space. As mentioned, the reason can be seen in the lack of a dimensional parameter. In the present paper, we demonstrate a simple example where we have such a parameter in empty (and flat) space. We consider a real scalar field $\phi$ in (1+1) dimensions with the action
\eq{7a}{ S&= \int dt\,dx\  \left[
	\frac12 \phi(t,x)\left(\pa_t^2-\pa_x^2+m^2\right)\phi(t,x)+\frac{\al}{4}\phi^4(t,x)\right],
}
obeying, after Fourier transform in time,  a non-linear Schr\"{o}dinger equation (NLS),
\eq{7}{ \left( -\om^2-\pa_x^2+\al \phi(x)^2 \right)\phi(x) &= 0.
}
The coupling $\al$ has the dimension  {\small \it length}$^{-2}$. As known, in the perturbative approach this model is super-renormalizable.

We start with a finite box with Dirichlet boundary conditions,
\eq{8}{ \phi(0)=\phi(L)=0,
}
resulting in discrete frequencies $\om_j$, and we calculate the vacuum (Casimir) energy generalizing \Ref{6}. Continuing,  it turns out that the limit $L\to\infty$ can be performed in a meaningful way  delivering the vacuum energy of a field with self-interaction in flat, empty space. This way, for the first time, the vacuum energy for a quantized field in flat, empty space is calculated. The key point is the self-interaction of the field, providing the necessary dimensional quantity.

It must be mentioned that the overwhelming majority of vacuum energy and Casimir force calculations were done for free fields with the 'only' complication of boundaries and/or background fields. Of course, for decades this was a serious problem. So, for example, it took 20 years to generalize the original calculation of Casimir \cite{casi48-51-793}, which was for flat, parallel plates, to a sphere \cite{boye68-174-1764}, and nearly 30 more years until the first calculation of vacuum energy in a three-dimensional background field \cite{bord96-53-5753}. For interacting fields, there are so far only perturbative calculations. Starting from a scalar field in a rectangular cavity, for example \cite{ford79-368-305} and  \cite{pete82-26-415}, over the electron field on a spherical boundary \cite{bord98-58-045003}, up to, recently, various improvements of the renormalization procedure \cite{valu18-133-401} and a very recent  perturbative calculation in \cite{song20}.We mention that also in \cite{dash75-12-2443} the quantum fluctuations were described by a free field.

From the point of view of quantum field theory we proceed as follows. The solutions $\phi_j(x)$ of the field equation \Ref{7} form a complete set and the expansion
\eq{8a}{ \phi(x) &= \sum_ja_j\phi_j(x)
}
holds with the corresponding eigenvalues $\om_j$. Further we apply the standard canonical quantization procedure turning the $a_j$ into operators with the usual commutation relations. As a result we arrive at the vacuum energy \Ref{4}. In short, we 'merely' substitute the usual plane waves by the functions $\phi_j(x)$.

In the present paper, we remind the basic formulas for the solutions of the NLS \Ref{7}. These were investigated in many papers and under many aspects, including the inverse scattering method. There is  rich and beautiful physics connected with these solutions, which we do not touch at all in this paper.
We use the explicit solutions in terms of Jacobi elliptic functions, which were investigated in the papers \cite{carr00-15-2645}, \cite{carr00-62-063610}, \cite{carr00-62-063611}, and also in the very recent \cite{sacc20-53-385204}. Subsequently, we remind some basic facts from the renormalization of vacuum energy. Finally, we calculate some examples and discuss the strong coupling behavior.

Throughout the paper, we use natural  units with $\hbar=c=1$. Further, we use the standard notations for the Jacobi elliptic functions as used, for instance, in \cite{AbramowitzStegun2010} with $m$ in place of the parameter $k^2$, such that $k^2=m$ and  $k'=\sqrt{1-m}$. Since $m$ is a standard notation in this context, we use the notation $\me$ for the mass (irrespective of whether 'e' denotes an electron or not).

\section{\label{T2}Solutions of the NLS in terms of elliptic functions}
As shown in  appendix B in \cite{seam05-71-033622},  any solution of \Ref{7} can be expressed in terms of one of the Jacobi elliptic functions. The specific choice is, then, motivated by the boundary conditions. It is meaningful to consider the cases $\al>0$ and $\al<0$ separately. We start with the repulsive one, i.e., with $\al>0$.

\subsection{\label{T2.1}The repulsive case}
We make the ansatz
\eq{2.1}{  \phi(x) &= A \ \sn (q x|m).
}
We mention that  $\sn(u|m)$ is that elliptic  function which generalizes $\sin(u)$.
The boundary condition \Ref{8} in $x=0$ is satisfied by the choice of the elliptic function $\sn$. The boundary condition in $x=L$ demands
\eq{2.2}{ q &= \frac{2j}{L}\,K(m),~~~(j=1,2,\dots),
}
where $K(m)$ is the complete elliptic integral of the first kind, which is a quarter period of $\sn$. $A$ and $m$ are still free parameters. We insert the ansatz into the equation \Ref{7} which we preliminarily rewrite in the form 
\eq{2.3}{ \phi''=-\om^2\phi+\al \phi^3.
}
Multiplying by $\phi$ we get with \Ref{2.2}  the first integral,
\eq{2.4}{q^2 \sn'^2 (q x|m) &= -\om^2 \sn^2 (q x|m)+\frac{\al}{2}\sn^4 (q x|m)+C,
}
where $C$ is an integration constant. With eq. (22.13.1) from \cite{AbramowitzStegun2010}, we arrive at
\eq{2.5}{ q^2(1-\sn)(1-m\, \sn (q x|m)) &= -\om^2 \sn^2 (q x|m)+\frac{\al}{2}\sn^4 (q x|m)+c.
}
Comparing equal powers of $\sn$, the relations
\eq{2.6}{ A^2=\frac{2}{\al}q^2m,~~~q^2(1+m)=\om^2,~~~q^2=c,
}
follow. Another relation emerges from the normalization condition which one must impose,
\eq{2.7}{ 1 &=\int_0^1dx\ \phi^2(x) ,
}
and which with \Ref{2.1} takes the form
\eq{2.8}{1 &= \frac{A^2}{q}\int_0^qdx\ \sn^2(x|m).
}
Here we apply eq. (5.134~1) from \cite{grad07} to get
\eq{2.9}{ 1=\frac{A^2}{qm}\left[q-E({\rm am}(q|m)|m)\right],
}
where ${\rm am}$ is the amplitude. It obeys
\eq{2.10}{ {\rm am}(2jK(m)|m)=j\pi,~~~E(j\pi|m)=2jE(m),
}
where $E(m)$ is the complete elliptic integral of second kind, and \Ref{2.8} turns into
\eq{2.11}{ 1 &= \frac{A^2}{K(m)\,m}(K(m)-E(m)).
}
Inserting from \Ref{2.6} and \Ref{2.2} delivers the equation
\eq{2.12}{ \frac{\al L^2}{(2j)^2} &= 2 K(m)(K(m)-E(m)).
}
This is an equation for the parameter $m$ and we denote the solutions by $ m_j$. Then the frequencies $\om$ become also discrete and from \Ref{2.6} we find
\eq{2.13}{   \om_j &= \frac{2j}{L}K(m_j)\sqrt{1+m_j}.
}
The calculation of these frequencies is quite easy since the right sides of \Ref{2.12} and \Ref{2.13} are monotone functions.

The $\om_j$, \Ref{2.13}, are the eigenfrequencies of the field $\phi(x)$  obeying the boundary conditions \Ref{8} and enter the vacuum energy \Ref{4}.
Below we will need their behavior  for large $j$. Expanding the mentioned right sides in powers of $m$, one comes to the perturbative (in $\al$)  expansion
\eq{2.14}{ \om_j  \underset{j \to \infty }{\sim}
	\frac{\pi j}{L}\left(1+\frac{3\al L^2}{4\pi^2 j^2}+\dots\right).
}
The leading order gives just the well-known eigenfrequencies of a free field in a box. The behavior for large $\al$ follows from \Ref{2.12} with the remark that the elliptic integrals grow logarithmically for $m\to1$. Using the corresponding asymptotic expansions,
\eq{2.14a}{ \om_j \underset{\al\to\infty}{\sim}\frac{\al L}{2j}
+\dots
}
follows for fixed $j$.

\subsection{\label{T2.2}The attractive case}
In distinction from the preceding subsection, we assume now $\al<0$ (we do not substitute $\al\to-\al$). Following \cite{carr00-62-063611}, we make the ansatz
\eq{2.15}{ \phi(x) &= A\ \cn(qx+\delta|m) ,
}
with the elliptic function $\cn$ which generalizes the cosine function. To fulfill the boundary condition \Ref{8} one must take the choice
\eq{2.16}{ q&= 2j K(m),~~~\delta =-K(m).
}
Further, the discussion goes completely in parallel to the preceding case. In place of \Ref{2.4} we have now
\eq{2.17}{ q^2 \cn'^2 &= -\om^2 \cn^2+\frac{\al}{2}\cn^4+c.
}
With eq. (22.13.2) from \cite{AbramowitzStegun2010} we arrive at
\eq{2.18}{ q^2(1-\cn)(1-m+m\, \cn) &= -\om^2 \cn^2+\frac{\al}{2}\cn^4+c,
}
and comparing equal powers of $\cn$ results in the relations
\eq{2.19}{ A^2 &=- \frac{2}{\al}q^2m,~~~q^2(1-2m)=\om^2,~~~q^2(1-m)=c.
}
Here one can understand the choice of the ansatz \Ref{2.15} since negative $\al$ in \Ref{2.6} would give an imaginary prefactor $A$. For the integration in the normalization condition, we  use now eq. (5.134~2) from \cite{grad07},
\eq{2.20}{ 1 &= \frac{A^2}{q \, m}
	\left(
E({\rm am}(q+\delta|m)|m)+(m-1)(q+\delta)
		-E({\rm am}(\delta|m)|m)-(m-1)\delta
	\right),
}
and with \Ref{2.16} we arrive at
\eq{2.21}{ 1 &= \frac{A^2}{K(m)m} (E(m)+(m-1)K(m)).
}
Inserting for $A$ from \Ref{2.19} delivers the equation
\eq{2.22}{ \frac{\al L^2}{(2j)^2} &=  2K(m)(E(m)+(m-1)K(m)),
}
whose solutions are the  $m_j$'s in the case  $\al<0$ (we do not introduce a separate notation). Finally, from \Ref{2.19}, the eigenfrequencies follow to be
\eq{2.23}{ \om_j &= \frac{2j}{L}K(m_j)\sqrt{1-2m_j}.
}
For large numbers $j$, the same expansion \Ref{2.14} as in the repulsive case holds. The behavior for large $\al$ is different from the preceding case. While, again,    from the right side of  \Ref{2.22} $m\to1$ follows, the frequencies $\om_j$ become imaginary as soon as $m$ exceeds $\frac12$. The critical value is
\eq{2.24}{  \frac{|\al| L^2}{(2j)^2}_{\big|\rm crit.} &\simeq
	1.57.
}
This means that the attractive self-interaction makes the system unstable if the coupling exceeds some critical value, which is, of course, an expected feature.

\section{\label{T3}The vacuum energy in a finite box}
The vacuum energy is, in zeta functional regularization, given by eq. \Ref{4}. In the case of a massive field, obeying the equation
\
\eq{3.1}{\left( -\om^2-\pa_x^2+\me^2+\al \phi(x)^2 \right)\phi(x) &= 0,
}
as we consider it here, it  reads
\
\eq{3.2}{ E_0 (s) &= \frac{\mu^{2s}}{2}\sum_{j=1}^\infty
	\left(\om_j^2+\me^2\right)^{\frac12-s},
}
with the eigenfrequencies $\om_j$ as given by \Ref{2.13} for $\al>0$ and by \Ref{2.23} for $\al<0$. In the latter case, when $\om_j$ is imaginary, still a stable solution exists as long as $\om_j^2\le \me^2$ holds. Imaginary $\om_j$ act similar to bound states in this context. Denoting the left side of \Ref{2.22} by $t$, this equation defines a function $m(t)$. Then the critical $t_{\rm cr.}$ is solution of the equation
\eq{3.3}{ t^{-2}K(m(t))^2(1-2m(t)) &= \frac{\me^2}{-\al}
}
and the condition of stationarity reads
\eq{3.4}{ \frac{-\al L^2}{(2j)^2}\le t_{\rm cr.}.
}
For $\me=0$ we come to \Ref{2.24}.

The renormalization procedure is, in general, well-known.  We follow the mentioned chapter in \cite{BKMM}, and formulas adopted to the (1+1)-dimensional case in the recent \cite{bord2012.14301}. First, we need to define the zeta function associated with our spectrum $\om_j$. It reads
\eq{3.5}{ \zeta_P(s) &= \sum_{j=1}^\infty \om_j^{-2s}.
}
The pole  parts, which we need for the heat kernel coefficients, follow from inserting the asymptotic expansion \Ref{2.14},
\eq{3.6}{\zeta_P(s) &= \sum_{j=1}^\infty
	  \left(\frac{\pi j}{L}\right)^{-2s}  \left( 1
	-2s\frac{3\al L^2}{4\pi^2}j^{-2-2s}+\dots\right)
\\\nn	&= \left(\frac{\pi}{L}\right)^{-2s}
	\left(      \zeta_{\rm R}(2s)
	-2s \frac{3\al L^2}{4\pi^2}\zeta_{\rm R}(2s+2)+\dots \ \right),
}
where, as already mentioned,  $\zeta_{\rm R}$ is the Riemann zeta function. The heat kernel coefficients are expressed by the residua,
\eq{3.7}{ a_n &={\rm Res}_{s=\frac12-n}\sqrt{4\pi}\,\Gamma(s)\zeta_P(s),
}
and those relevant for the renormalization read
\eq{3.8}{ a_0&=L,~~~a_\frac12 =-\sqrt{\pi},~~~a_1=-\frac32\al L.
}
As usual, $a_0$ is the volume of the empty space. The coefficient $a_\frac12$ is, may be occasionally, the same as in \cite{bord2012.14301} and we refer to the remark in the paragraph following eq. (81) in that paper. The coefficient $a_1$ is the deciding one for the renormalization. It is non-zero with the known consequences. In passing we mention, that it is zero for $\al=0$ which explains the finite result \Ref{6}.

The renormalization is  performed by subtracting the contributions from the heat kernel coefficients \Ref{3.8} to $E_0(s)$. Using the (1+1)-dimensional modification of eq. (4.30) in \cite{BKMM}, we define
\eq{3.9}{ E_0^{div}(s) &=
	\left(\frac{1}{8\pi s}
	-\frac{1}{4\pi}
	\left(\frac12+\ln\frac{\me}{2\mu}\right)\right) \me^2\,a_0
	+\frac{1}{4\sqrt{\pi}}\, \me\,a_\frac12
\\\nn&~~~~	+\left(
	\frac{-1}{8\pi s}+\frac{1}{4\pi}\left(1+\ln\frac{\me}{2\mu}\right)
	\right)a_1.
}
As known, the heat kernel expansion provides an asymptotic expansion of the vacuum energy in inverse powers of the mass and \Ref{3.9} collects all non-decreasing powers. We define  the renormalized vacuum energy by
\eq{3.10}{ E_0^{ren} &= \lim_{s\to0}\left(E_0(s)-E_0^{div}(s)\right),
}
such that it vanishes for $\me\to\infty$. This is a kind of normalization condition based on the argument that a 'heavy' quantum field should not have vacuum fluctuations. At once, this way the regularization ambiguity, which came in with the parameter $\mu$, drops out (see below, eq. \Ref{3.16}).

We mention that the poles in $s$ are the only ultraviolet divergences in the considered model. For instance, these are at most first order in the coupling $\al$ in agreement with the superrenormalizability of the model.

In order to perform the continuation $s\to0$ in \Ref{3.10}, we use the asymptotic expansion \Ref{2.14} to split the regularized vacuum energy into two parts,
\eq{3.11}{E_0(s) &= E_0^{lead}(s)+E_0^{sub},
}
with a 'leading' contribution,
\eq{3.12}{ E_0^{lead}(s) &=\frac{\mu^{2s}}{2} \sum_{j=1}^\infty
	\left[ \left(\frac{\pi}{L}\right)^{1-2s}
	+(1-2s)\frac{3\al+2\me^2}{4}\left(\frac{\pi}{L}\right)^{-1-2s}
	\right],
\\\nn  &= \frac{\mu^{2s}}{2} \left(\frac{\pi}{L}\right)^{1-2s}
		\left[\zeta_{\rm R}(2s-1)+(1-2s)
		\frac{(3\al+2\me^2)L^2}{4\pi^2} \zeta_{\rm R}(2s+1)
		\right],
}
and
\eq{3.13}{E_0^{sub} &=\frac12\sum_{j=1}^\infty
	\left[ \left(\om_j^2+\me^2\right)^{\frac12}
	-\left(\frac{\pi j}{L}+\frac{(3\al+2\me^2)L}{4\pi j}
	\right) \right].
}
In the last expression we could put $s=0$ since the sum converges. Note, $\om_j$ is to be calculated from \Ref{2.13} for $\al>0$ and from \Ref{2.23} for $\al<0$. Numerical evaluation does not pose problems.

\begin{figure}[h]
	\centering
	\includegraphics[width=0.7\linewidth]{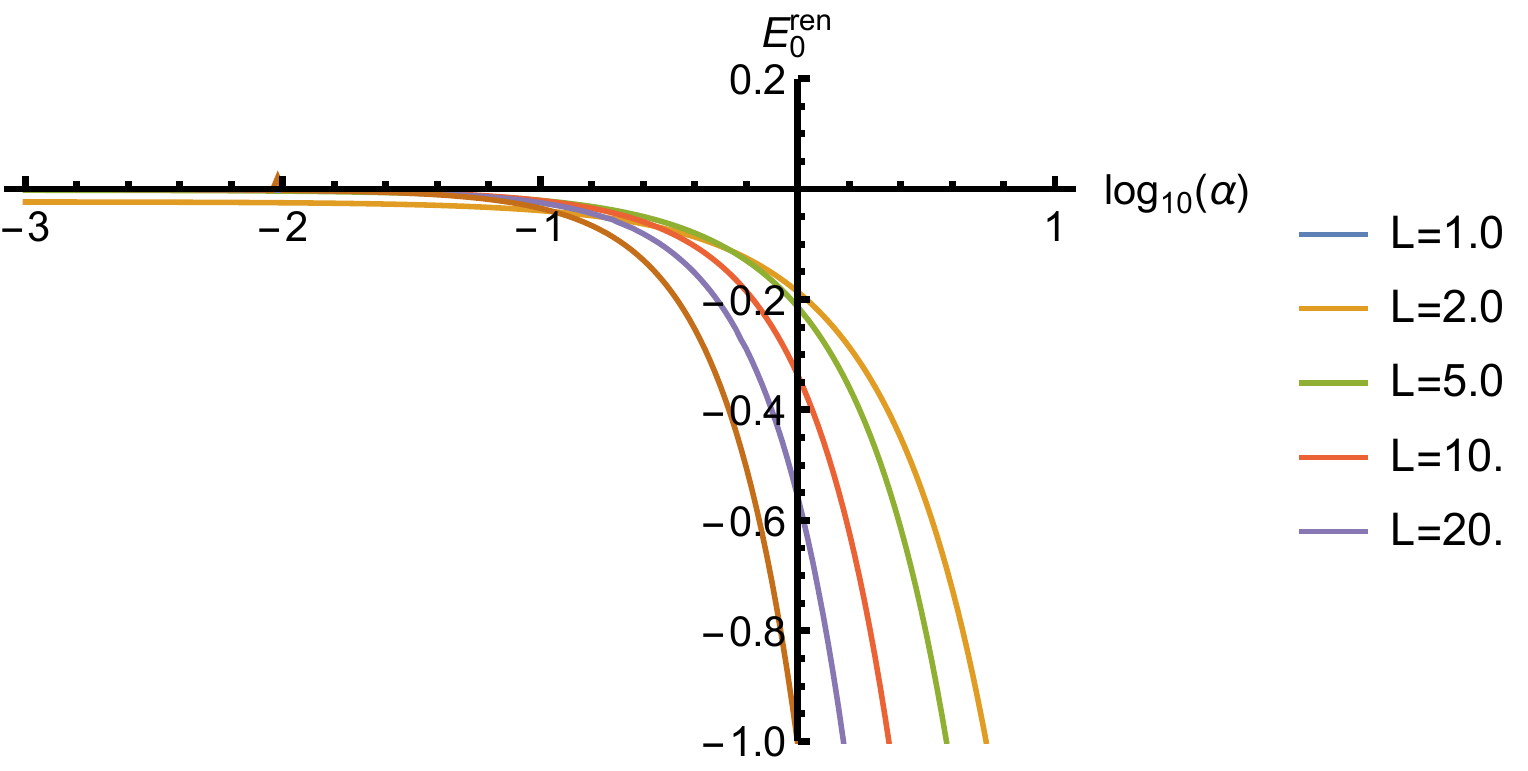}
	\caption[Spectrum]{The vacuum energy of a massive scalar field in a box  with repulsive self-interaction with $\me=1$.}
	\label{fig:1}
\end{figure}

Rearranging \Ref{3.10} and using \Ref{3.11}, we arrive at
\eq{3.14}{ E_0^{ren}  &= E_0^{as}+E_0^{sub}
}
with
\eq{3.15}{ E_0^{as} &= \lim_{s\to0}\left(E_0^{lead}(s)-E_0^{div}(s) \right).
}
Inserting \Ref{3.9} into \Ref{3.12} we get
\eq{3.16}{ E_0^{as} &=
	\frac{1}{8\pi L}
	\left[ -\me^2L^2+2\pi\me L-\frac{\pi^2}{3}
	+(3\al+2\me^2)L^2\left(\ga+\ln\frac{\me L}{2\pi}\right)\right].
}
Representation \Ref{3.14} holds for both signs of the coupling $\al$. It can be evaluated numerically. Examples are shown in Fig. \ref{fig:1}. For $\al\to 0 $, together with $\me\to0$,  the free case \Ref{6} is recovered. The same holds for $L\to 0$.

\begin{figure}[h]
	\centering
	\includegraphics[width=0.7\linewidth]{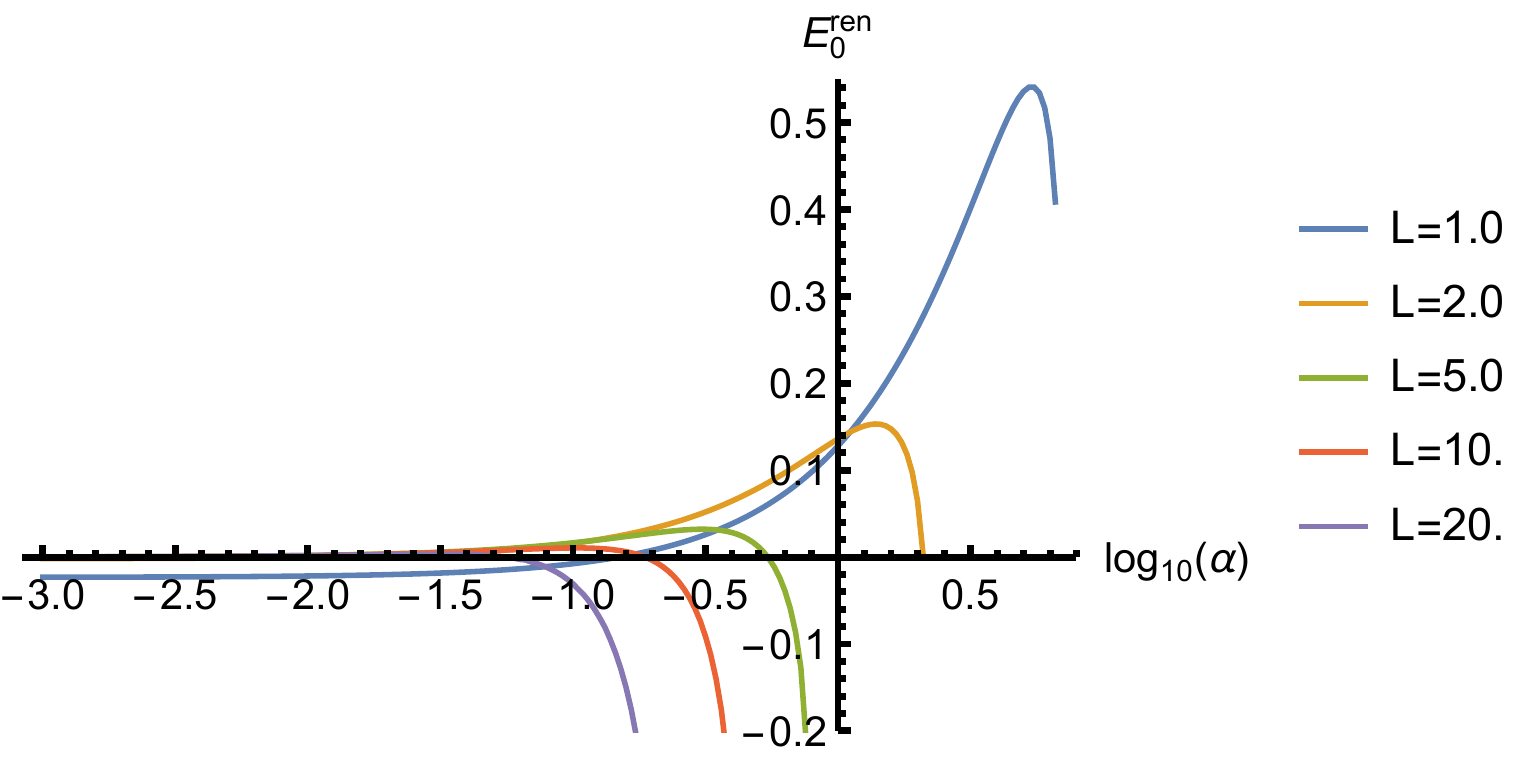}
	\caption[Spectrum]{The vacuum energy of a massive scalar field in a box with attractive self-interaction with $\me=1$. The curves terminate where the energy becomes complex.}
	\label{fig:2}
\end{figure}

For the massless case, the renormalization procedure is slightly different. As known, and as can be seen from \Ref{3.16}, the limit $\me\to0$ is logarithmically divergent. In the massless case, the argument with the large mass expansion is not applicable and without further considerations, the result will depend on   $\mu$, thus it will be not unique. In this paper, we do not discuss possible normalization conditions in the massless case and keep the $\mu$-dependence of the result.

For $s\to0$, the vacuum energy $E_0(s)$ has a pole term resulting from $a_1$ in \Ref{3.9} or, equivalently, from $\zeta_{\rm R}(2s+1)$ in \Ref{3.12}. It reads
\eq{3.17}{ E_0(s) &= \frac{3\al L}{16\pi s}+O(1).
}
For the renormalization we simply subtract this pole contribution and define, in analogy to \Ref{3.15},
\eq{3.17a}{E_0^{ren\ (m=0)} &=\lim_{s\to 0} \left(E_0^{(m=0)}(s)-\frac{3\al L}{16\pi s}\right).
}
Also, we split, in parallel to \Ref{3.11}
\eq{3.17b}{ E_0^{(m=0)}(s) &= E_0^{lead\ (m=0)}(s)+E_0^{sub\ (m=0)},
}
where $E_0^{lead\ (m=0)}(s)$ is given by \Ref{3.12} and $E_0^{sub\ (m=0)}$ by \Ref{3.13}, both with $\me=0$.
Rearranging, we come to
\eq{3.18}{ E_0^{ren~(m=0)} &= E_0^{as~(m=0)}+ E_0^{sub},
}
with
\eq{3.19}{  E_0^{as~(m=0)} &=
	 \lim_{s\to0} \left(E_0^{lead\ (m=0)}(s) -\frac{3\al}{16\pi s} \right)
\\\nn
	&= \frac{1}{8\pi L}\left[ 3\al L^2\left(-1+\ga+\ln\frac{\mu L}{\pi}\right)
		-\frac{\pi^2}{3}\right]
}
($\ga$ is Euler's gamma). Examples are shown in Fig. \ref{fig:2}.

\begin{figure}[h]
	\centering
	\includegraphics[width=0.7\linewidth]{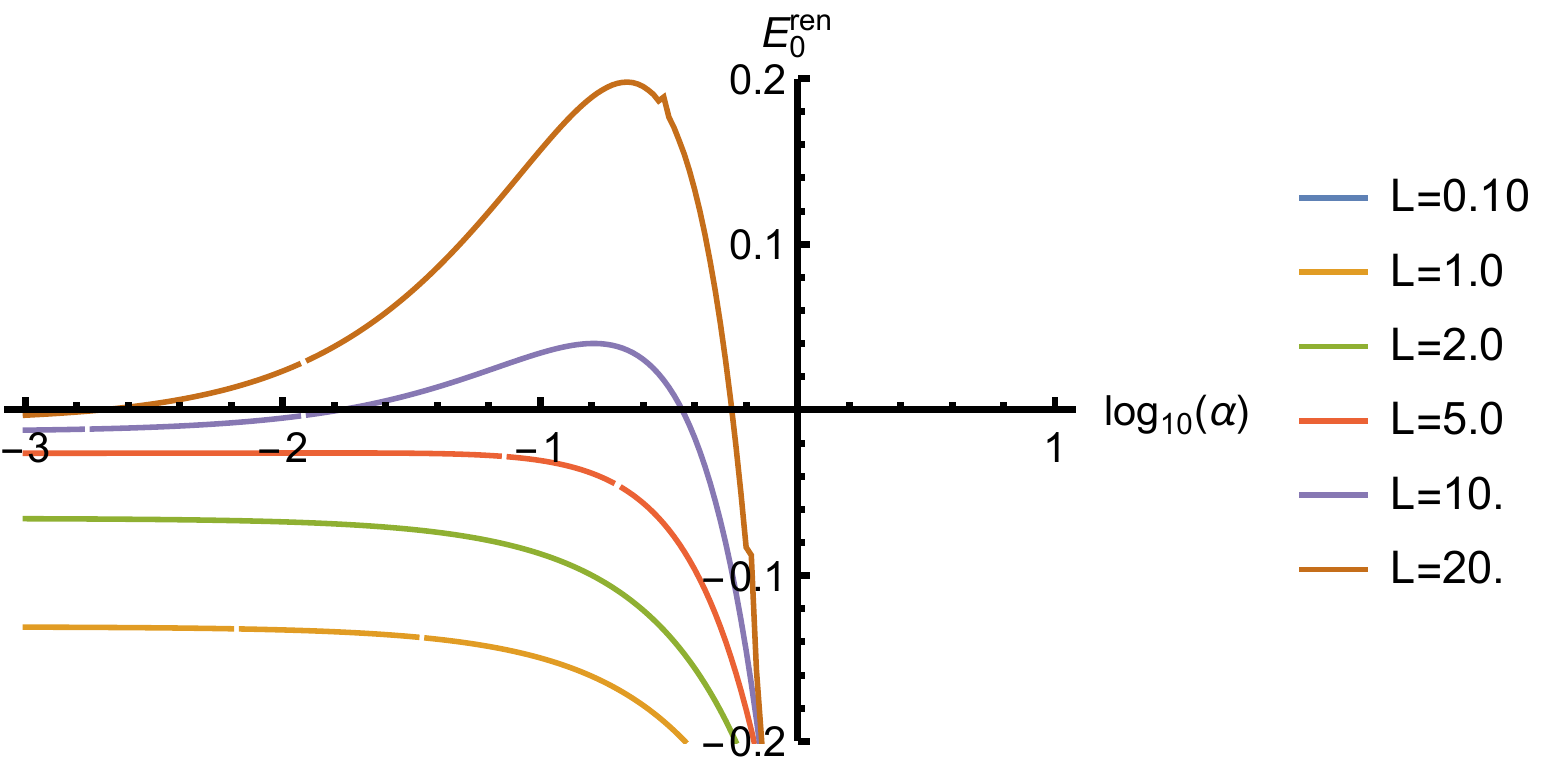}
	\caption[Spectrum]{The vacuum energy of a massless scalar field in a box with repulsive self-interaction and $\mu=1$.}
	\label{fig:3}
\end{figure}

\begin{figure}[h]
	\centering
	\includegraphics[width=0.7\linewidth]{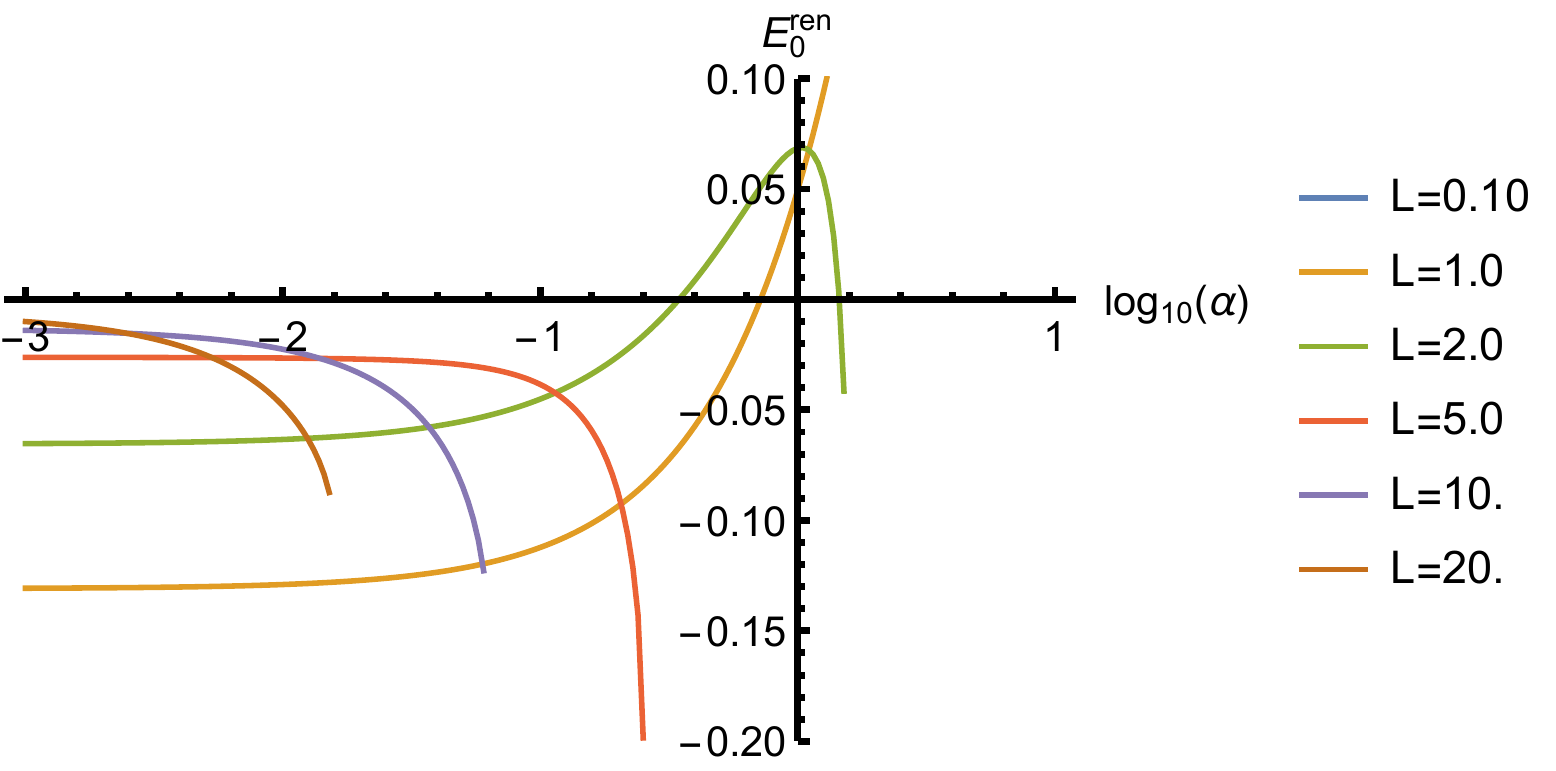}
	\caption[Spectrum]{The vacuum energy of a massless scalar field in a box  with attractive self-interaction and $\mu=1$.  The curves terminate when the energy becomes complex.}
	\label{fig:4}
\end{figure}

\section{\label{T4}The infinite volume case}
By taking the limit $L\to\infty$ we come to the infinite volume case. In this limit, first of all, we mention that for $\al<0$ we always have an imaginary frequency  since  \Ref{3.4} cannot be fulfilled for any finite $\al$ and $j$. So we restrict yourselves now to $\al>0$. In the limit, the frequencies $\om_j$, \Ref{2.13} become continuous and the sum in the regularized vacuum energy \Ref{3.2} becomes an integral, according to
\eq{4.1}{ \frac{1}{L}\sum_{j=1}^\infty \underset{L\to\infty}{\to}\int_0^\infty dq,
	~~~\frac{j}{L}\underset{L\to\infty}{\to}q.
}
Thereby the vacuum energy becomes proportional to $L$. Since in this limit, the homogeneity of the space is restored, the energy density, $\frac{1}{L}E_0(s)$,  is a finite constant and it is just this quantity we are interested in. Not to introduce another symbol, we keep the notation $\frac{1}{L}E_0(s)$ also in the limit.
Thus, the vacuum energy is now given by
\eq{4.2}{\frac{1}{L}E_0(s) &= \frac{\mu^{2s}}{2}\int_0^\infty dq \
	\left\{\left[
	2q \, g\left(\frac{\al}{4q^2}\right)\right]^2
	+\me^2\right\}^{\frac12-s},
}
where the function $g(t)$ is defined as follows. We rewrite equation \Ref{2.12} in the form
\eq{4.3}{ t &= 2K(m)(K(m)-E(m)),
}
with
\eq{4.4}{  t  &=\frac{\al}{4q^2}.
}
Equation \Ref{4.3} defines a function $m(t)$ (this is the same function as used in the beginning of Sect. 3 in a different context). Then the expression \Ref{2.13} for the  frequency turns into
\eq{4.5}{ \om=2qK(m(t))\sqrt{1+m(t)}\equiv 2q\, g(t),
}
where we introduced the new notation $g(t)$. This is a smooth function of $t$ with the asymptotics
\eq{4.6}{ g(t)\underset{t\to0}{=} \frac{\pi}{2}\left(1+\frac{3t}{\pi^2}+\dots\right),
	~~~ g(t)\underset{t\to\infty}{=}\frac{1+\sqrt{1+2t}}{\sqrt{2}}+\dots .
}
The asymptotics for $t\to0$ is the same as in \Ref{2.14} and that for $t\to\infty$ can be obtained from expanding eqs. \Ref{4.3} and \Ref{4.5} for $m\to 1$. It is up to exponentially small contributions. Eq. \Ref{2.14a} is a special case of it.

It is convenient to change the integration in \Ref{4.2} from $q$ to $t=\frac{\al}{4q^2}$ to get the representation
\eq{4.7}{ \frac{1}{L}E_0(s) &= \frac{\mu^{2s}\al^{1-s}}{2}
	\int_0^\infty dt\, t^{s-2}\left( g^2(t)+\frac{t\me^2}{\al}\right)^{\frac12-s}.
}
Now the ultraviolet divergence sits in $t=0$. Using the corresponding zeta function, which is related to the vacuum energy in zeta functional regularization by
\eq{4.8}{ \zeta_P(s) &= 2 \frac{1}{L}E_0(s+1/2) ,
}
and the same methods as in Sect. 3, we get the heat kernel coefficients
\eq{4.9}{ a_0&= 1,~~~a_\frac12=-\sqrt{\al\pi},~~~a_1=-\frac32\al.
}
It must be mentioned that $a_0$ and $a_1$ appear from those in finite volume, eq. \Ref{3.8}, as density, i.e., simply divided by $L$. In distinction, $a_{\frac12}$ from  \Ref{3.8} would vanish after division by $L$, whereas  $a_{\frac{1}{2}}$ from  \Ref{4.9} is non-zero. As a consequence, the limits of infinite volume ($L\to\infty$) and removing the regularization ($s\to0$) do not commute.

\begin{figure}[h]
	\centering
	\includegraphics[width=0.7\linewidth]{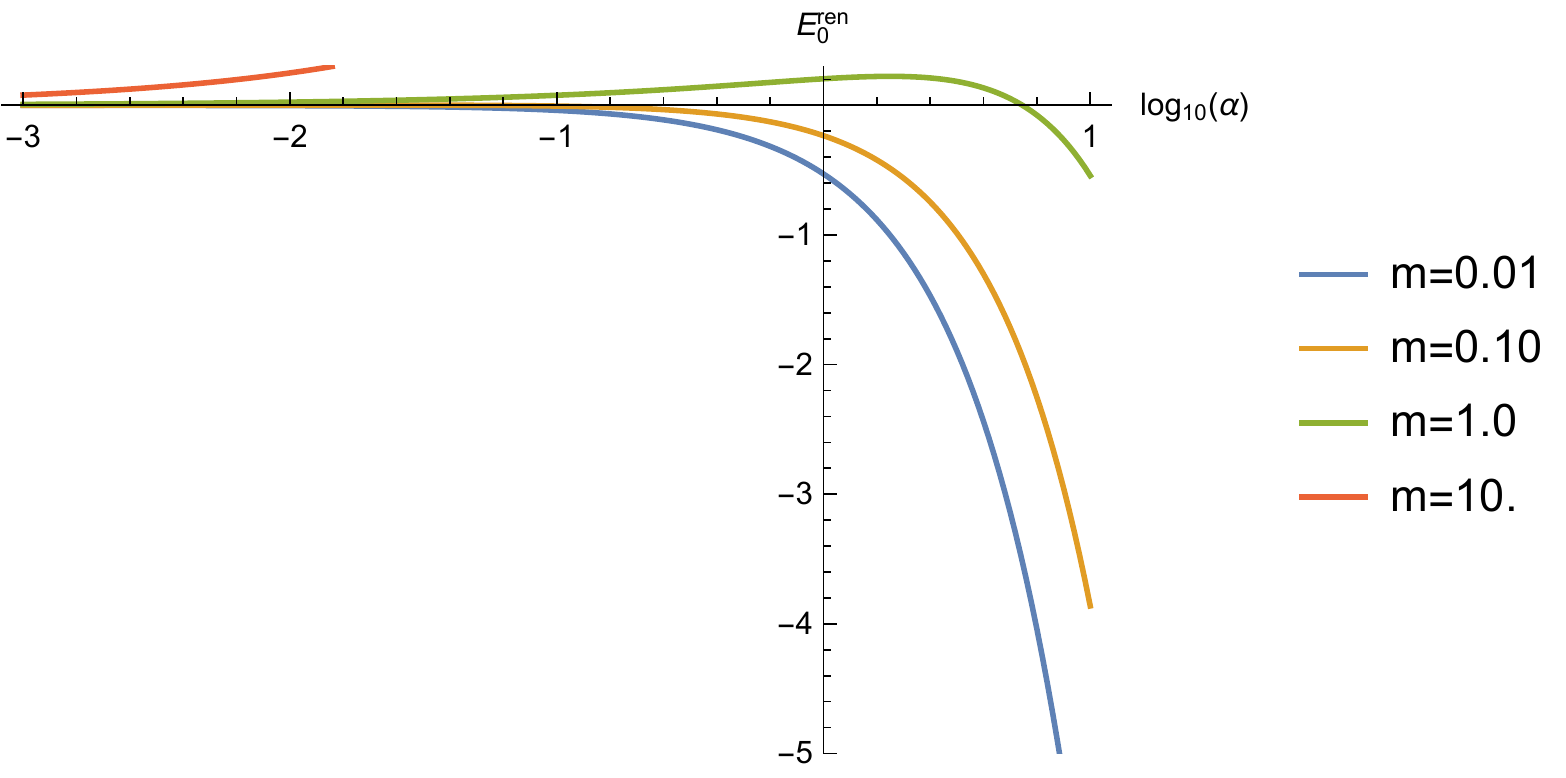}
	\caption[Spectrum]{The vacuum energy of a massive scalar field with repulsive self-interaction on the whole axis.  }
	\label{fig:5}
\end{figure}

For the renormalization in the massive case we use the same methods as above. For instance, we define $\frac{1}{L}E_0^{div}$ by eq. \Ref{3.9} with the coefficients \Ref{4.9}. Further, we use eqs. \Ref{3.10}, divided by $L$,
\eq{4.10}{\frac{1}{L}E_0^{ren} &= \lim_{s\to0}
	\left(\frac{1}{L}E_0(s) -\frac{1}{L}E_0^{div}(s) \right).
}
However, for technical reasons, we change the splitting and in place of \Ref{3.11} we use
\eq{4.11}{ \frac{1}{L}E_0(s) &= \frac{1}{L}E_0^{lead}(s)
	+\frac{1}{L}E_0^{sub}
}
with
\eq{4.12}{\frac{1}{L}E_0^{lead}(s) &=
	\frac{\mu^{2s}\al^{1-s}}{2}
	\int_0^1 dt\, t^{s-2}
	\left(\frac{\pi}{2}\right)^{\frac12-s}
	\left[1+(1-2s)\frac{(3\al+2\me^2)t}{2\pi^2\al}\right]^{\frac12-s}
}
for the 'leading' part and
\eq{4.13}{\frac{1}{L}E_0^{sub} &= E_A+E_B
}
with %
\eq{4.14}{ E_A &= \frac{\al}{8}\int_0^1dt\, t^{-2}
		\left[
		\left( g^2(t)+\frac{t\me^2}{\al}\right)^{\frac12}
		-\frac{\pi}{2}\left(1+\frac{(3\al+2\me^2)t}{2\pi\al}\right)
		\right],
\\\nn	 E_B &= \frac{\al}{8}\int_1^\infty dt\, t^{-2}
			\left( g^2(t)+\frac{t\me^2}{\al}\right)^{\frac12}.
}
The integrals in \Ref{4.14} converge and we could put $s=0$. The asymptotic part is now defined, similar to \Ref{3.15}, by
\eq{4.15}{ \frac{1}{L}E_0^{as} &=\lim_{s\to0}
		\left(
		\frac{1}{L}E_0^{lead}(s)-\frac{1}{L}E_0^{div}(s)
		\right)
}
Inserting and taking the limit results in
\eq{4.16}{\frac{1}{L}E_0^{as} &=
	\frac{1}{8\pi}\left(-(3\al+2\me^2)\ln\frac{\sqrt{\al}\pi}{\me}
	-\me^2+2\pi\sqrt{\al}\,\me-\frac{\al\pi^2}{2}\right).
}
This way, by eqs. \Ref{4.10}, \Ref{4.14} and \Ref{4.16}, we have formulas for numerically computing the vacuum energy in the considered case. Examples are shown in Fig. \ref{fig:5}. We mention that for numerical reasons the integration in $E_B$ was split,
\eq{4.17}{E_B=E_{B1}+E_{B2}
}
with
\eq{4.18}{ E_{B1} &=  \frac{\al}{8}\int_1^{t_p} dt \, t^{-2}
	\left( g^2(t)+\frac{t\me^2}{\al}\right)^{\frac{1}{2}},
\\\nn E_{B2} &= \frac{\al}{8}\int_{t_p}^\infty dt\, t^{-2}
			\left[\left(\frac{1+\sqrt{1+2t}}{\sqrt{2}}\right)^2
			+\frac{t\me^2}{\al}\right]^{\frac{1}{2}},
}
where above the division point $t_p$ the asymptotic expression from \Ref{4.6} is used. In the numerical work a value of $t_p=100$ was sufficient.

\begin{figure}[h]
	\centering
	\includegraphics[width=0.7\linewidth]{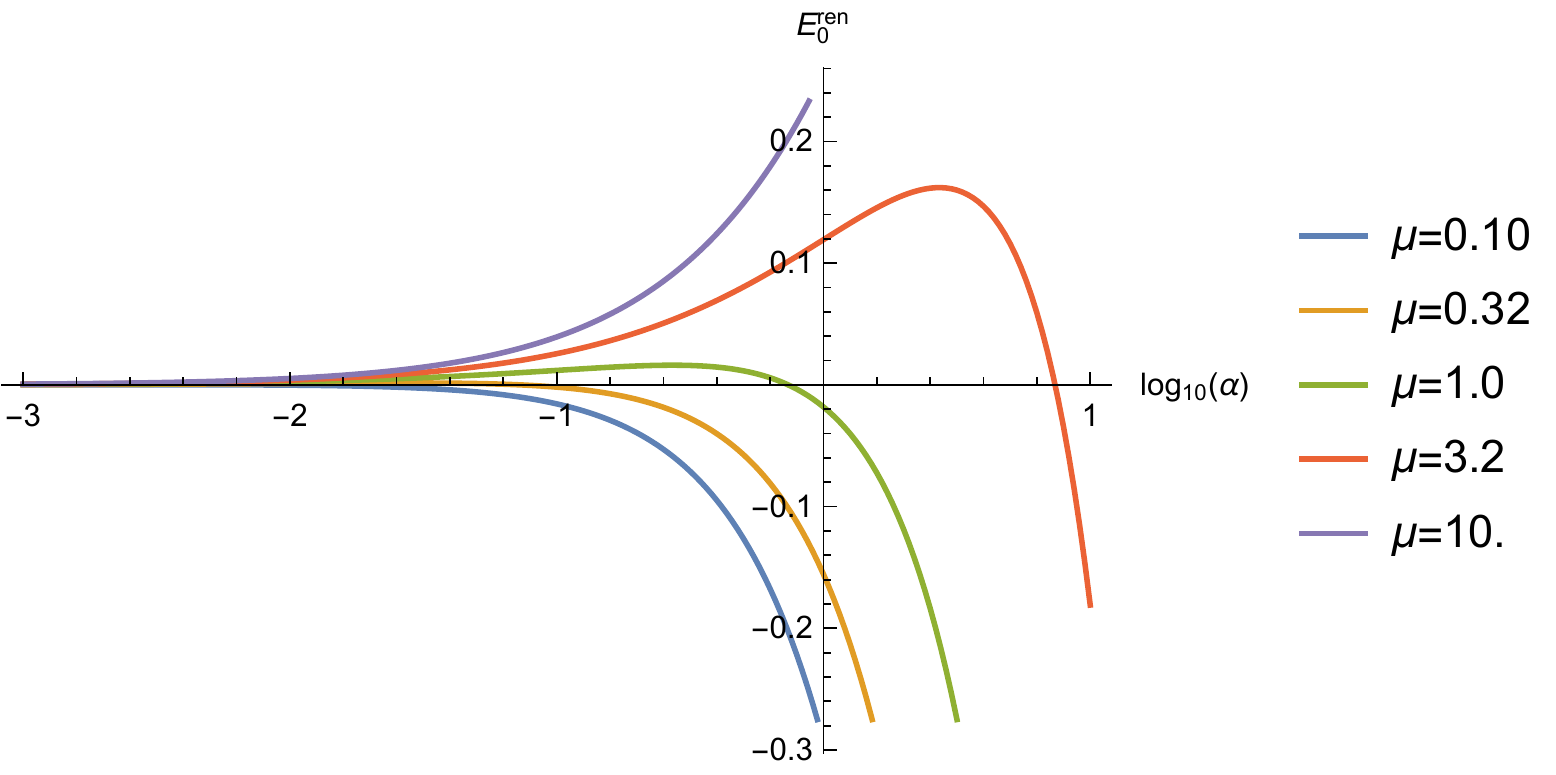}
	\caption[Spectrum]{The vacuum energy of a massless scalar field with repulsive self-interaction on the whole axis.  }
	\label{fig:6}
\end{figure}

In the massless case the same remarks as in Sect. \ref{T3} apply. The vacuum energy is
\eq{4.19}{\frac{1}{L}E_0(s) &= \frac{\mu^{2s}\al^{1-s}}{8}
	\int_0^\infty dt \, t^{s-2}g^{1-2s}(t).
}
The pole term for $s\to0$ can be obtained either from the heat kernel coefficient $a_1$, \Ref{4.9}, and \Ref{3.9}, or from inserting the asymptotic expansion \Ref{4.6} and it reads
\eq{4.20}{ \frac{1}{L}E_0(s) &=\frac{3\al}{16\pi s}+O(1).
}
Again, in the massless case we define the renormalized vacuum energy by subtracting the pole term,
\eq{4.21}{ \frac{1}{L}E_0^{ren} &=
	\lim_{s\to0}\left(\frac{1}{L}E_0(s)-\frac{3\al}{16\pi s}\right).
}
Further we split
\eq{4.22}{ \frac{1}{L}E_0(s) &=\frac{1}{L}E_0^{lead}(s)
	+\frac{1}{L}E_0^{sub}
}
with, from the asymptotic expansion \Ref{4.6},
\eq{4.23}{ \frac{1}{L}E_0^{lead}(s) &=
	\frac{\mu^{2s}\al^{1-s} }{8}
	\int_0^1 dt \, t^{s-2}
	\left(\frac{\pi}{2}\right)^{1-2s}
	\left[1+(1-2s)\frac{3 t}{\pi^2}\right],
}
and the 'subtracted' part,
\eq{4.24}{ \frac{1}{L}E_0^{sub} &=
	\frac{\al}{8}\int_0^1 dt\, t^{-2}
	\left[ g(t)-
	\frac{\pi}{2}\left(1+\frac{3 t}{\pi^2}\right)    \right]
	+	\frac{\al}{8}\int_1^\infty dt\, t^{-2}g(t),
}
where we could put $s=0$. Finally, rearranging contributions gives us
\eq{4.25}{ \frac{1}{L}E_0^{ren} &= \frac{1}{L}E_0^{as}-
	\frac{1}{L}E_0^{sub}
}
with
\eq{4.26}{   \frac{1}{L}E_0^{as} &= \lim_{s\to0}
	\left( \frac{1}{L}E_0^{lead}-\frac{3\al}{16\pi s}\right),
\\\nn
	&= \frac{3\al}{8\pi}\left(-\ln\frac{\sqrt{\al}\pi}{2\mu}-1-\frac{\pi^2}{6}\right)}
This vacuum energy can now be calculated numerically. For technical reasons also the same splitting of $E_0^{sub} $ as in eq. \Ref{4.18} was used. Examples are shown in Fig. \ref{fig:6}.

\section{\label{T5}The limit of strong coupling}
In this section we consider the limit $\al\to\infty$, i.e., the limit of strong coupling in eqs. \Ref{7} and \Ref{3.1}, of the renormalized vacuum energy, which was calculated in the preceding two sections.

We start with the massive cases and equip the relevant energies with an index '1' in place of the index '0' which mentions that this is a vacuum energy. For finite $L$, we have the energy \Ref{3.14} with its constituent parts \Ref{3.16} and \Ref{3.13}. The subtracted energy \Ref{3.13} can be written in the form
\eq{5.1}{ E_1^{sub} &= \frac{\sqrt{\al}}{2}\sum_{j=1}^\infty
	\left\{
	\left[\left(\frac{2j}{\sqrt{\al}L}g\left(\frac{\al L}{4j^2}\right)\right)^2
	+\frac{\me^2}{\al}\right]^{\frac12}
	-  \left[\frac{\pi j}{\sqrt{\al}L}
	+\frac{3}{4\pi}\left(1+\frac23\frac{\me^2}{\al}\right)
	\frac{\sqrt{\al}L}{j}\right]                  \right\}.
}
Here, the function $g(t)$ is the same as defined by \Ref{4.5}. Before we can consider the limit $\al\to\infty$, we need to rewrite the last term in the square bracket,
\eq{5.2}{	\frac{\sqrt{\al}L}{j}  \to
	\frac{1}{\frac{j}{\sqrt{\la}L}+1}  + \frac{1}{\frac{j}{\sqrt{\la}L}}-\frac{1}{\frac{j}{\sqrt{\la}L}+1}.
}
We get two terms. In the second,
\eq{5.3}{\tilde{E}_1 &\equiv \frac{\sqrt{\al}}{2}\frac{-3}{4\pi}
	\left(1+\frac23\frac{\me^2}{\al}\right)
	\sum_{j=1}^\infty   \left(
	\frac{1}{\frac{j}{\sqrt{\la}L}}-\frac{1}{\frac{j}{\sqrt{\la}L}+1}
	\right),
}
we carry out the summation,
\eq{5.4}{\tilde{E}_1 & = \frac{-3\al L}{8\pi}\left(1+\frac23\frac{\me^2}{\al}\right)H(\sqrt{\al}L),
}
where $H(n)$ is the harmonic number. In the limit we get
\eq{5.5}{\tilde{E}_1 & \underset{\al\to\infty}{\simeq}  \frac{-3\al L}{8\pi}(\ga+\ln(\sqrt{\al}L)).
}
In the first term, for $\al\to\infty$, the summation over $j$ turns into an integration (similar to \Ref{4.1}) with
\eq{5.6}{ \sum_{j=1}^\infty & \to \sqrt{\al} L\int_0^\infty dq,~~~\frac{j}{\sqrt{\al}L} \to q
}
and we come to
\eq{5.7}{ {E}^{sub}_1 & \underset{\al\to\infty}{\simeq}
	\frac{\al L}{2}\int_0^\infty dq\, \left[2q\, g\left(\frac{1}{2q}\right)-\left(\pi q+\frac{3}{4\pi}\frac{1}{q+1}\right)\right]    +\tilde{E}_1 .
}
Here, the integral gives a number and we introduce the notation
\eq{5.8}{A &\equiv  \frac{4\pi}{3}\int_0^\infty dq\,
	\left[2q\, g\left(\frac{1}{2q}\right)
	-\left(\pi q+\frac{3}{4\pi}\frac{1}{q+1}\right)\right] ,
\\\nn	&=  \frac{\pi}{3}\int_0^\infty \frac{dt}{t^2}
		 \left[g(t)-\left(\frac{\pi}{2}+\frac{3}{2\pi}\frac{t}{1+2\sqrt{t}}\right)\right],
\\\nn	&\simeq 1.9932.
}
In the second line we made the substitution $q=\frac{1}{2\sqrt{t}}$. This way, we get from  \Ref{5.7}
\eq{5.9}{ {E}^{sub}_1 & \underset{\al\to\infty}{\simeq}
	\frac{3\al L}{8\pi} \left(-\ln({\sqrt{\al}L})-\ga+A \right).
}
Together with \Ref{3.19} we get from \Ref{3.14}
\eq{5.10}{ E_1^{ren}   & \underset{\al\to\infty}{\simeq}
	\frac{3\al L}{8\pi} \left(-\ln\frac{\sqrt{\al}\,2\pi}{\me}+A \right),
}
which is up to terms of zeroth order in $\al$.

Next we consider the massive case in infinite volume. We give it the index '3'. The asymptotic part of the energy is given by eq. \Ref{4.16} and we note
\eq{5.11}{ \frac{1}{L}E_3^{as} &  \underset{\al\to\infty}{\simeq}
	\frac{3\al}{8\pi}\left(-\ln\frac{\sqrt{\al}\, \pi}{\me}-\frac{\pi^2}{2}\right).
}
The subtracted part of the energy is given by eq. \Ref{4.13} and we note
\eq{5.12}{  \frac{1}{L}E_3^{sub} &  \underset{\al\to\infty}{\simeq}
	\frac{\al}{8}\left\{
	\int_0^1 \frac{dt}{t^{2}}  \left[
	g(t)-\left(\frac{\pi}{2}+\frac{3t}{2\pi}\right)  \right]
+\int_1^\infty    \frac{dt}{t^{2}}  g(t)
\right\}.
}
The integrals over $t$ can be brought into the form of the second line in \Ref{5.8},
\eq{5.13}{  \frac{1}{L}E_3^{sub} &  \underset{\al\to\infty}{\simeq}
		\frac{3\al}{8\pi}\left(A+\frac{\pi^2}{6}-\ln  2\right),
}
and we arrive at
\eq{5.14}{  \frac{1}{L}E_3^{ren} &  \underset{\al\to\infty}{\simeq}
		\frac{3\al }{8\pi} \left(-\ln\frac{\sqrt{\al}\,2\pi}{\me}+A \right).
}
This is the same as \Ref{5.10}.	
	
We turn to the massless case and consider  finite volume. This case gets the index '2'. The asymptotic part of the vacuum energy is given by \Ref{3.19} and takes in the limit the form
\eq{5.15}{ E_2^{as} &  \underset{\al\to\infty}{\simeq}
	\frac{3\al L}{8\pi}\left(\ln\frac{\mu L}{\pi}-1+\ga\right).
}
The subtracted part, \Ref{3.13},  is, in this limit, the same as $E_1^{sub}$, \Ref{5.9}, and for the renormalized energy we arrive
\eq{5.16}{  E_2^{ren} &  \underset{\al\to\infty}{\simeq}
		\frac{3\al L}{8\pi}\left(-\ln\frac{\sqrt{\al}\pi}{\mu}+A-1\right).
}
Finally, we have the case 4 which is infinite volume and massless field. The asymptotic part is given by eq. \Ref{4.26} and reads in the limit
\eq{5.17}{  \frac{1}{L}E_4^{as} &  \underset{\al\to\infty}{\simeq}
	\frac{3\al}{8\pi} \left(-\ln\frac{\sqrt{\al}\pi}{2\mu}-1-\frac{\pi^2}{6}\right).
}
The subtracted part, \Ref{4.24},  is the same as in case 3 and reads
\eq{5.18}{  \frac{1}{L}E_4^{sub} &  \underset{\al\to\infty}{\simeq}
		\frac{3\al}{8\pi}\left(A+\frac{\pi^2}{6}-\ln 2\right) .
}
Taking these two together we arrive at
\eq{5.19}{ \frac{1}{L}E_4^{ren} &  \underset{\al\to\infty}{\simeq}
		\frac{3\al }{8\pi} \left(-\ln\frac{\sqrt{\al}\,\pi}{\mu}+A-1 \right)
}
which is the same as \Ref{5.16}, case 2.

As we have seen, the cases 1 and 3, as well as 2 and 4, have the same limit, which shows that this limit is independent on the volume. The massless cases, which depend on the arbitrary parameter $\mu$, give the same as the massive cases with the formal substitution $\mu\to \frac{e}{2}\me$.

On the leading logarithmic behavior of the renormalized vacuum energy  a more general view is possible. Let us consider the regularized vacuum energy, \Ref{3.2}, for large $\al$. We represent the $\om_j$, \Ref{2.13}, in the form
\eq{5.20}{  \om_j &=\frac{2j}{L}\, g\left(\frac{\al L^2}{4j^2}\right),
}
where we used the function $g(t)$, which was defined by \Ref{4.5}. The regularized vacuum energy becomes
\eq{5.21}{ E_0(s) &= \frac{\mu^{2s}\al^{\frac{1}{2}-s}}{2}
	\sum_{j=1}^\infty \left[\left(
	\frac{2j}{\sqrt{\al}L}\, g\left(\frac{\al L^2}{4j^2}\right) \right)^2
	+\frac{\me^2}{\al}  \right]^{\frac{1}{2}-s},
}
which is similar to \Ref{5.1}. Now we consider the limit $\al\to\infty$. The sum turns into an integration, following \Ref{5.6}. The mass term disappears and we arrive at
\eq{5.22}{  E_0(s) &= \frac{\al L}{2}\left(\frac{\mu^2}{\al}\right)^s
	\int_0^\infty dt\,t^{s-2} g^{1-2s}(t),
}
where we made the substitution $q=1/2\sqrt{t}$ alike in \Ref{5.8}. The integral is a function only of $s$. It has a pole in $s=0$ and the pole term is, of course, the same as in \Ref{3.9} (for $\me=0$). This pole term generates the logarithmic contribution,
\eq{5.23}{ E_0(s) &= \frac{-a_1}{8\pi s}\left(1-s\ln\frac{\al}{\mu^2}+\dots\right),
}
such that after subtracting the pole term, i.e., after preforming the renormalization, we get
\eq{5.24}{ E_0^{ren} &= \frac{a_1}{8\pi}\ln\al +\dots\,.
}
The dots denote the terms without $\ln\al$ and depend on the renormalization scheme. Accounting for $a_1$, \Ref{3.9} or \Ref{4.9}, we reproduce the leading order in the above mentioned examples, \Ref{5.10}, \Ref{5.14}, \Ref{5.16} and \Ref{5.19}.

\section{\label{T6}Conclusions}
We have calculated the vacuum energy for a scalar field with $\phi^4$ self-interaction in $(1+1)$-dimensions in a box of length $L$ with Dirichlet boundary conditions and in the whole space for both, massive and massless fields. For dimensional reasons, the vacuum energy has the following structure. We use the definition of the cases given in Sect. \ref{T5},
\begin{equation}\label{6.1} \begin{array}{rclll}
	\hspace{-3cm}1.~~~~~~	 E_1^{ren} &=& \frac{1}{L}f_1\left(\frac{\me}{\sqrt{\al}},\me L\right), ~~~~~~~ &(L\mbox{ finite}),~~&\me\ne 0,
	\\[5pt]\nn
	\hspace{-3cm}2.~~~~~~	 E_2^{ren} &=& \frac{1}{L}f_2\left(\frac{\mu}{\sqrt{\al}},\mu L\right),  &(L\mbox{ finite}),~~&\me= 0,
	\\[5pt]\nn
	\hspace{-3cm}3.~~~~~~	 E_3^{ren} &=& \frac{1}{L}f_3\left(\frac{\me}{\sqrt{\al}} \right),  &(L\mbox{ infinite}),~~&\me\ne 0,
	\\[5pt]\nn
	\hspace{-3cm}4.~~~~~~	 E_4^{ren} &=& \frac{1}{L}f_4\left(\frac{\mu}{\sqrt{\al}} \right),  &(L\mbox{ infinite}),~~&\me= 0,
\end{array}\end{equation}
where the functions $f_i$ are dimensionless. These are shown in the figures in Sects. \ref{T3} and \ref{T4}. We remind that for the infinite volume cases we consider the energy density.

The case 4 carries little new information. The vacuum energy \Ref{5.19} has  only the factor in front of the logarithm as sensible information since the constant term remains undefined because of the arbitrary parameter $\mu$. Further, the mentioned factor follows, for dimensional reasons, from the heat kernel coefficient $a_1$, \Ref{4.9}.

In the massive case, the renormalization is done by subtracting the contributions from the relevant heat kernel coefficients, which delivers a unique result. In distinction, in the massless case the arbitrary parameter $\mu$ remains present in the renormalized energy.

This way, we were able to attribute a sensible vacuum energy to a field with self-interaction in a flat and empty space. We underline that this result is not perturbative but calculated in all orders of the self-interaction.

A characteristic feature is the strong coupling behavior. It is calculated in Sect. \ref{T5} and shows in all four cases the same result of a growing  negative vacuum energy. This allows for the interpretation that the vacuum energy makes the system unstable with respect to an increasing coupling. For smaller coupling, positive values are possible as can be seen from the figures. For vanishing coupling we reobtain the known result for a free field.

As only relation to a physical system we mention the relation $\al=\la^{-2}$ to the dimensionless parameter $\lambda$ introduced in \cite{carr00-62-063610}, whose value in   BEC  experiments is $\la\sim1/25$, implying $\al\sim600$.

A final remark concerns the possible embedding of the vacuum energy calculated in this paper into some context. Obvious candidates would be the already mentioned BEC, where eq. \Ref{7} has the meaning of the Gross-Pitaevski equation, or an effective potential in the sense of Coleman-Weinberg. Since we were in the present paper only interested in the first calculation of vacuum energy for an interacting field, these questions are left for future work.

\bibliographystyle{unsrt}
\bibliography{C:/Users/bordag/WORK/Literatur/bib/papers}

\end{document}